\documentclass{JHEP}
\usepackage{epsfig}

\newcommand{\eref}[1]{(\ref{#1})}

\newcommand{\fref}[1]{Figure~\ref{#1}}
\newcommand{\cref}[1]{Chapter~\ref{#1}}
\newcommand{\beq}{\begin{equation}}
\newcommand{\eeq}{\end{equation}}
\newcommand{\ba}{\begin{array}}
\newcommand{\ea}{\end{array}}
\newcommand{\bcenter}{\begin{center}}
\newcommand{\ecenter}{\end{center}}

\def\IB{\relax\hbox{$\inbar\kern-.3em{\rm B}$}}
\def\IC{\relax\hbox{$\inbar\kern-.3em{\rm C}$}}
\def\ID{\relax\hbox{$\inbar\kern-.3em{\rm D}$}}
\def\IE{\relax\hbox{$\inbar\kern-.3em{\rm E}$}}
\def\IF{\relax\hbox{$\inbar\kern-.3em{\rm F}$}}
\def\IG{\relax\hbox{$\inbar\kern-.3em{\rm G}$}}
\def\IGa{\relax\hbox{${\rm I}\kern-.18em\Gamma$}}
\def\IH{\relax{\rm I\kern-.18em H}}
\def\IK{\relax{\rm I\kern-.18em K}}
\def\IL{\relax{\rm I\kern-.18em L}}
\def\IP{\relax{\rm I\kern-.18em P}}
\def\IR{\relax{\rm I\kern-.18em R}}
\def\IZ{\relax\ifmmode\mathchoice
{\hbox{\cmss Z\kern-.4em Z}}{\hbox{\cmss Z\kern-.4em Z}}
{\lower.9pt\hbox{\cmsss Z\kern-.4em Z}}
{\lower1.2pt\hbox{\cmsss Z\kern-.4em Z}}\else{\cmss Z\kern-.4em Z}\fi}
\def\II{\relax{\rm I\kern-.18em I}}

\def\sCC{{\kern 0.27em\vrule height1.45ex width0.03em depth0em
          \kern-0.30em\rm C}}
\def\C{{\mathchoice
  {\sCC}
  {\sCC}
  {\kern 0.225em \vrule height1.05ex width0.025em depth0em \kern-0.25em \rm C}
  {\kern 0.180em \vrule height0.78ex width0.02em depth0em \kern-0.2em \rm C}
        }}
\def\sHH{{\rm I\kern-.16em{}H}}
\def\H{{\mathchoice
  {\sHH}
  {\sHH}
  {\rm I\kern-.13em{}H}
  {\rm I\kern-.13em{}H} }}
\def\sNN{{\rm I\kern-.16em{}N}}
\def\N{{\mathchoice
  {\sNN}
  {\sNN}
  {\rm I\kern-.12em{}N}
  {\rm I\kern-.10em{}N} }}
\def\sPP{{\rm I\kern-.16em{}P}}
\def\P{{\mathchoice
  {\sPP}
  {\sPP}
  {\rm I\kern-.12em{}P}
  {\rm I\kern-.10em{}P} }}
\def\sQQ{{\kern 0.27em \vrule height1.45ex width0.03em depth0em
          \kern-0.30em \rm Q}}
\def\Q{{\mathchoice
        {\sQQ}
        {\sQQ}
  {\kern 0.225em \vrule height1.05ex width0.025em depth0em \kern-0.25em \rm Q}
  {\kern 0.180em \vrule height0.78ex width0.020em depth0em \kern-0.20em \rm Q}
        }}
\def\sRR{{\rm I\kern-0.16em{}R}}
\def\R{{\mathchoice
  {\sRR}
  {\sRR}
  {\rm I\kern-0.12em{}R}
  {\rm I\kern-0.10em{}R} }}
\def\sZZ{{\rm Z\kern-0.32em{}Z}}
\def\Z{{\mathchoice
  {\sZZ}
  {\sZZ} 
  {\rm Z\kern-0.3em{}Z}     
  {\rm Z\kern-0.25em{}Z} }}  
\def\ZZZ{{\rm Z\kern-0.24em{}Z}}
\def\sII{{\rm I\kern-0.16em{}I}}
\def\I{{\mathchoice
  {\sII}
  {\sII}
  {\rm I\kern-0.12em{}I}
  {\rm I\kern-0.10em{}I} }}

\def\Hom{{\rm Hom}}

\def\inbar{\,\vrule height1.5ex width.4pt depth0pt}
\font\cmss=cmss10 \font\cmsss=cmss10 at 7pt

\def\odd{{\rm odd}}
\def\even{{\rm even}}
\def\rtimes{\mbox{$\times\!\rule{0.3pt}{1.1ex}\,$}}

\def\smiley{\hbox{\large$\bigcirc$\hspace{-0.80em}\raise.2ex
\hbox{$\cdot\cdot$}\kern-.61em\lower.2ex\hbox{\scriptsize$\smile$}}\ }
\def\frowny{\hbox{\large$\bigcirc$\hspace{-0.80em}\raise.2ex
\hbox{$\cdot\cdot$}\kern-.635em\lower.2ex\hbox{\scriptsize$\frown$}}\ }

\def\I{{\rlap{1} \hskip 1.6pt \hbox{1}}}

\newcommand{\gen}[1]{\langle #1 \rangle}
\newcommand{\mat}[1]{\left( \matrix{#1} \right)}
\makeatletter
\let\hangafter\@hangfrom
\makeatother

\newtheorem{definition}{\sf DEFINITION}

\newtheorem{lemma}{\sf LEMMA}

\newtheorem{theorem}{\sf THEOREM}

\newtheorem{proposition}{\sf PROPOSITION}

\newtheorem{corollary}{\sf COROLLARY}

\preprint{MIT-CTP-3026\\ \\ {\tt hep-th/0010023}}
\title{Discrete Torsion, Non-Abelian Orbifolds and the Schur Multiplier}
\author{Bo Feng, Amihay Hanany, Yang-Hui He and Nikolaos Prezas
\footnote{
Research supported in part
by the Reed Fund Award, 
the CTP and the LNS of MIT and the U.S. Department of Energy 
under cooperative research agreement \# DE-FC02-94ER40818.
A. H. is also supported by an A. P. Sloan Foundation Fellowship,
and a DOE OJI award.}
\\
Center for Theoretical Physics,
\\ Massachusetts Institute of Technology,\\ Cambridge, MA 02139, USA.\\
\email{fengb, hanany, yhe, prezas@ctp.mit.edu}
}
\abstract{Armed with the explicit computation of Schur
Multipliers, we offer a classification of $SU(n)$ orbifolds for
$n=2,3,4$ which
permit the turning on of discrete torsion. This is
in response to the host of activity lately in vogue on the
application of discrete torsion to D-brane orbifold theories. As a
by-product, we find a hitherto unknown class of ${\cal N}=1$ 
orbifolds with non-cyclic discrete torsion group.
Furthermore,
we supplement the {\it status quo ante} by investigating a first example of
a non-Abelian orbifold admitting discrete torsion, namely the ordinary
dihedral group as a subgroup of $SU(3)$. A comparison of the quiver
theory thereof with that of its
covering group, the binary dihedral group, without discrete torsion,
is also performed.}
\begin{document}
\newpage
\section{Introduction}
The study of string theory in non-trivial NS-NS B-field backgrounds has of
late become one of the most pursued directions of research. Ever since
the landmark papers \cite{SW}, where it was shown
that in the presence of such non-trivial B-fields along the
world-volume 
directions of the D-brane, the gauge theory living thereupon
assumes a non-commutative guise in the large-B-limit, most works were
done in this direction of space-time non-commutativity. However, there
is an alternative approach in the investigation of the effects of the
B-field, namely {\bf discrete torsion}, which is of great
interest in this respect. On the other hand, as discrete torsion
presents itself to be a natural generalisation to the study of orbifold
projections of D-brane probes at space-time singularities, a topic
under much research over the past few years, it is also
mathematically and physically worthy of pursuit under this light.

A brief review of the development of the matter from a historical
perspective shall serve to guide the reader. Discrete torsion first
appeared in \cite{Vafa} in the study of the
closed string partition function $Z(q,\bar{q})$ on the orbifold $G$. 
And shortly thereafter, it effects on the geometry of space-time were
pointed out \cite{VafaWit}.
In particular, \cite{Vafa} noticed that
$Z(q,\bar{q})$ could contain therein, phases $\epsilon(g,h) \in U(1)$
for $g,h \in G$, coming from the twisted sectors of the theory, as
long as
\beq\label{epsilon}
\ba{l}
\epsilon(g_1g_2,g_3) = \epsilon(g_1,g_3)\epsilon(g_2,g_3)\\
\epsilon(g,h) = 1/\epsilon(h,g)\\
\epsilon(g,g) = 1,
\ea
\eeq
so as to ensure modular invariance.

Reviving interests along this line, Douglas and Fiol
\cite{Doug,DougFiol} extended discrete torsion to the open string
sector by showing that the usual procedure of projection by orbifolds
on D-brane probes \cite{DM,LNV}, applied to {\bf projective
representations} instead of the ordinary {\em linear representations}
of the orbifold group $G$, gives exactly the gauge theory with discrete
torsion turned on. In other words, for the invariant matter fields which
survive the orbifold, $\Phi$ such that $\gamma^{-1}(g) \Phi \gamma(g)
= r(g) \Phi,\quad\forall~g \in G$, we now need the representation
\beq\label{proj1}
\ba{l}
\gamma(g) \gamma(h) = \alpha(g,h) \gamma(gh),\quad g,h \in G~~$with$\\
\alpha(x,y)\alpha(xy,z) = \alpha(x,yz)\alpha(y,z), \qquad
        \alpha(x,\II_G) = 1 = \alpha(\II_G,x) \quad
        \forall x,y,z \in G, \qquad \\
\ea
\eeq
where $\alpha(g,h)$ is known as a cocycle. These cocycles constitute,
up to the equivalence
\beq\label{proj2}
\alpha(g,h) \sim \frac{c(g)c(h)}{c(gh)} \alpha(g,h),
\eeq
the so-called second cohomology group $H^2(G,U(1))$
of $G$, where $c$ is any function (not necessarily a homomorphism)
mapping $G$ to $U(1)$; this is what we usually mean by 
{\em discrete torsion being classified by $H^2(G,U(1))$}.
We shall formalise all these definitions in the subsequent sections.

In fact, one can show \cite{Vafa}, that the choice
\[
\epsilon(g,h) = \frac{\alpha(g,h)}{\alpha(h,g)},
\]
for $\alpha$ obeying \eref{proj1} actually satisfies \eref{epsilon}, whereby
linking the concepts of discrete torsion in the closed and open string
sectors.
We point this out as one could be easily confused as to the
precise parametre called discrete torsion and which is actually
classified by the second group cohomology.

Along the line of \cite{Doug,DougFiol}, a series of papers by
Berenstein, Leigh and Jejjala \cite{BL,BJL} developed the technique
to study the {\em non-commutative} moduli space of the ${\cal N}=1$ 
gauge theory living on $\IC^3/\IZ_m \times \IZ_n$ parametrised as
an algebraic variety. A host of activities followed in the
generalisation of this abelian orbifold, notably to $\C^4/\IZ_2
\times \IZ_2 \times \IZ_2$ by \cite{Ray}, to the inclusion of
orientifolds by \cite{Klein}, and to the orbifolded conifold by
\cite{Tatar}.

Along the mathematical thread, Sharpe has presented a prolific series
of works to relate discrete torsion with connection on gerbes
\cite{Sharpe}, which could allow generalisations of the concept to
beyond the 2-form B-field. Moreover, in relation to twisted K-theory
and attempts to unify space-time cohomology with group cohomology in
the vein of the McKay Correspondence (see e.g. \cite{HeSong}), works 
by Gomis \cite{Gomis} and Aspinwall-Plesser \cite{AspinPles,Aspin}
have given some guiding light.

Before we end this review of the current studies, we would like to
mention the work by Gaberdiel \cite{Gab}. He pointed out that there
exists a different choice, such that the original intimate relationship
between discrete torsion in the closed string sector and the
non-trivial cocycle in the open sector can be loosened. It would be
interesting to investigate further in this spirit.

We see however, that during these last three years of renewed
activity, the focus has mainly been on Abelian orbifolds. It is one of
the main intentions of this paper to initiate the study of non-Abelian
orbifolds with discrete torsion, which, to the best of our knowledge,
have not been discussed so far in the literature\footnote{In the
context of conformal field theory on orbifolds, there has been a 
recent work addressing some non-Abelian cases \cite{Katrin}.}. 
We shall classify
the general orbifold theories with ${\cal N}=0,1,2$ supersymmetry which
{\em could allow discrete torsion} by exhaustively computing the
second cohomology of the discrete subgroups of $SU(n)$ for $n=4,3,2$.

Thus rests the current state of affairs. Our main objectives are
two-fold: to both supplement the past, by presenting and studying
a first example of a non-Abelian orbifold which affords discrete
torsion, and to presage the future, by classifying the orbifold
theories which could allow discrete torsion being turned on.
\newpage
\section*{Nomenclature}
Throughout this paper, unless otherwise specified, we shall adhere to
the following conventions for notation:\\
\begin{tabular}{ll}
$\omega_n$ & $n$-th root of unity;\\
$G$ & finite group of order $|G|$; \\
$\IF$ & (algebraically closed) number field; \\
$\IF^*$ & multiplicative subgroup of $\IF$; \\
$\gen{x_i|y_j}$ & the group generated by elements $\{x_i\}$ with
	relations $y_j$;\\ 
$<G_1,G_2,\ldots,G_n>$ & group generated by the generators of groups
$G_1,G_2,\ldots,G_n$; \\
$\gcd(m,n)$ & the greatest common divisor of $m$ and $n$; \\
$D_{2n},E_{6,7,8}$ & ordinary dihedral, tetrahedral, octahedral and
	icosahedral groups;\\ 
$\widehat{D_{2n}},\widehat{E_{6,7,8}}$ & 
	the binary counterparts of the above; \\
$A_n$ and $S_n$ & alternating and symmetric groups on $n$ elements; \\
$H \triangleleft G$ & $H$ is a normal subgroup of $G$; \\
$A \rtimes B$ & semi-direct product of $A$ and $B$;\\
$Z(G)$ & centre of  $G$;\\
$N_G(H)$ & the normaliser of $H \subset G$;\\
$G':=[G,G]$ & the derived (commutator) group of $G$;\\
$\exp(G)$ & exponent of group $G$.\\
\end{tabular}
\section{Some Mathematical Preliminaries}
\subsection{Projective Representations of Groups}
We begin by first formalising \eref{proj1}, the group representation
of our interest:
\begin{definition}
A {\bf projective} representation of $G$ over a field $\IF$ 
(throughout we let $\IF$ be an algebraically closed field with
 characteristic $p\geq 0$) is a
mapping $\rho : G \rightarrow GL(V)$ such that
\[
(A)~~~ \rho(x)\rho(y) = \alpha(x,y)\rho(xy)~~\forall~~x,y \in G;
\qquad
(B)~~~ \rho(\II_G) = \II_V.
\]
\end{definition}
Here $\alpha : G \times G \rightarrow \IF^*$ is a mapping whose
meaning we shall clarify later.
Of course we see that if $\alpha=1$ trivially, then we have our familiar
ordinary representation of $G$ to which we shall refer as {\em
linear}. Indeed, the mapping $\rho$ into $GL(V)$ defined above is
naturally equivalent to a homomorphism into the projective linear group
$PGL(V) \cong GL(V)/\IF^*\II_V$, and hence the name ``projective.''
In particular we shall be concerned with projective {\em
matrix} representations of $G$ where we take $GL(V)$ to be
$GL(n,\IF)$.

The function $\alpha$ can not be arbitrary and two immediate
restrictions can be placed thereupon purely from the structure of the group:
\beq
\label{alpha}
\ba{clcl}
(a)& \mbox{Group Associativity} & \Rightarrow &
	\alpha(x,y)\alpha(xy,z) = \alpha(x,yz)\alpha(y,z), \quad\forall
	x,y,z \in G\\
(b)& \mbox{Group Identity} & \Rightarrow &
	\alpha(x,\II_G) = 1 = \alpha(\II_G,x), \quad \forall x \in G.
\ea
\eeq
These conditions on $\alpha$ naturally leads to another discipline of
mathematics.
\subsection{Group Cohomology and the Schur Multiplier}
The study of such functions on a group satisfying \eref{alpha}
is precisely the subject of the
theory of {\bf Group Cohomology}. In general we let $\alpha$ to take
values in $A$, an abelian coefficient group ($\IF^*$ is certainly a
simple example of such an $A$) 
and call them {\bf cocycles}. The set of all cocycles we
shall name $Z^2(G,A)$. Indeed it is straight-forward to see that
$Z^2(G,A)$ is an abelian group. We subsequently define a set of
functions satisfying
\beq
\label{cobound}
B^2(G,A) := \{(\delta g)(x,y) := g(x) g(y) g(xy)^{-1}\}
\quad\mbox{for any~}g :G \rightarrow A \mbox{~such that~} g(\II_G)=1,
\eeq
and call them {\em coboundaries}.
It is then obvious that $B^2(G,A)$ is a (normal) subgroup of
$Z^2(G,A)$ and in fact constitutes an equivalence relation on the
latter in the manner of \eref{proj2}. Thus it becomes 
a routine exercise in cohomology to define
\[
H^2(G,A) := Z^2(G,A) / B^2(G,A),
\]
the {\em second cohomology} group of $G$.

Summarising what we have so far, we see that the projective
representations of $G$ are
classified by its second cohomology $H^2(G,\IF^*)$. To facilitate the
computation thereof, we shall come to an important concept:
\begin{definition}
The {\bf Schur Multiplier} $M(G)$ of the group $G$ is the second
cohomology group with respect to the trivial action of $G$ on $\IC^*$:
\[
M(G) := H^2(G,\IC^*).
\]
\end{definition}

Since we shall be mostly concerned with the field $\IF = \IC$, the
Schur multiplier is exactly what we need. However, the properties
thereof are more general. In fact, for any algebraically closed field
$\IF$ of zero characteristic,
$M(G)\cong H^2(G,\IF^*)$. In our case of $\IF =
\IC$, it can be shown that \cite{Klein},
\[
H^2(G,\IC^*) \cong H^2(G,U(1)).
\]
This terminology is the more frequently encountered one
in the physics literature.

One task is thus self-evident: the calculation of the Schur Multiplier
of a given group $G$ shall indicate possibilities of projective
representations of the said group, or in a physical language, the
possibilities of turning on discrete torsion in string theory on the
orbifold group $G$. In particular, if $M(G) \cong \II$, then the
second cohomology of $G$ is trivial and no non-trivial discrete
torsion is allowed. We summarise this
\[
\mbox{KEY POINT:}\quad \mbox{Calculate }M(G) \Rightarrow 
\mbox{Information on Discrete Torsion.}
\]
\subsection{The Covering Group}
The study of the actual projective representation of $G$ is very
involved and what is usually done in fact is to ``lift to an ordinary
representation.'' What this means is that for a central 
extension\footnote{i.e., $A$ in the centre
	$Z(G^*)$ and $G^*/A \cong G$ according to the exact sequence
$1 \rightarrow A \rightarrow G^* \rightarrow G \rightarrow 1$.} 
$A$ of $G$ to $G^*$, we say a projective
representation $\rho$ of $G$ {\bf lifts} to a linear
representation $\rho^*$ of $G^*$ if (i) $\rho^*(a \in A)$ is
proportional to $\II$ and (ii) there is a section\footnote{i.e., for the
	projection $f:G^*\rightarrow G$, $\mu \circ f = \II_G$.}
$\mu : G \rightarrow G^*$ such that $\rho(g) = \rho^*(\mu(g)),\quad
\forall g \in G$. Likewise it {\em lifts projectively} if $\rho(g) =
t(g) \rho^*(\mu(g))$ for a map $t:G \rightarrow \IF^*$. Now we are
ready to give the following:
\begin{definition}\label{defcover}
We call $G^*$ a {\bf covering group}\footnote{
	Sometimes is also known as {\bf representation group}.}
of $G$ over $\IF$ if the following are satisfied:\\
(i) $\exists$ a central extension $1 \rightarrow A \rightarrow G^*
\rightarrow G \rightarrow 1$ such that any projective representation
of $G$ lifts projectively to an ordinary representation of $G^*$;\\
(ii) $|A| = |H^2(G,\IF^*)|$.
\end{definition}
The following theorem, initially due to Schur, characterises covering groups.
\begin{theorem}\label{cover}{\rm (\cite{Karp} p143)}
$G^{\star}$ is a covering
group of $G$ over $\IF$ if and only if the following conditions hold:\\
(i) $G^{\star}$ has a finite subgroup $A$ with $A\subseteq
Z(G^{\star}) \cap [G^{\star},G^{\star}]$;\\
(ii) $G \cong G^{\star}/A$;\\
(iii) $|A|=|H^2(G,F^{\star})|$\\
where $[G^{\star},G^{\star}]$ is the derived group\footnote{For a
	group $G$, $G' := [G,G]$ is the group generated by elements of
	the form $xyx^{-1}y^{-1}$ for $x,y \in G$.}
$G^{*'}$ of $G^*$.
\end{theorem}

Thus concludes our prelude on the mathematical rudiments, the utility
of the above results shall present themselves in the ensuing.
%
\section{Schur Multipliers and String Theory Orbifolds}
The game is thus afoot. Orbifolds of the form $\IC^k/\{G \in SU(k)\}$
have been widely studied in the context of gauge theories living on
D-branes probing the singularities. We need only to compute $M(G)$ for
the discrete finite groups of $SU(n)$ for $n=2,3,4$ 
to know the discrete torsion
afforded by the said orbifold theories.
\subsection{The Schur Multiplier of the Discrete Subgroups of
$SU(2)$}
Let us first remind the reader of the well-known $ADE$
classification of the discrete finite subgroups of $SU(2)$.
Here are the presentations of these groups:
\beq
\label{SU2}
\ba{|c|c|c|c|}
\hline
G	& $Name$ & $Order$	& $Presentation$ \\
\hline
\widehat{A_n} & $Cyclic$, \cong \IZ_{n+1} & n & \gen{a|a^n=\II} \\ \hline
\widehat{D_{2n}} & \mbox{Binary Dihedral} & 4n &
		\gen{a,b|b^2=a^n, abab^{-1}=\II} \\ \hline
\widehat{E_6} & \mbox{Binary Tetrahedral} & 24 &
		\gen{a,b|a^3=b^3=(ab)^3} \\ \hline
\widehat{E_7} & \mbox{Binary Octahedral} & 48 &
		\gen{a,b|a^4=b^3=(ab)^2} \\ \hline
\widehat{E_8} & \mbox{Binary Icosahedral} & 120 &
		\gen{a,b|a^5=b^3=(ab)^2} \\ \hline
\ea
\eeq
We here present a powerful result due to Schur (1907) (q.v. Cor. 2.5,
Chap. 11 of \cite{Karp2}) which aids us to explicitly compute large
classes of Schur multipliers for finite groups:
\begin{theorem} {\rm (\cite{Karp} p383)}
\label{Schur}
Let $G$ be generated by $n$ elements with (minimally)
$r$ defining relations and let the Schur multiplier $M(G)$ have a
minimum of $s$ generators, then
\[
r \ge n+s.
\]
In particular, $r=n$ implies that $M(G)$ is trivial and $r=n+1$, that
$M(G)$ is cyclic.
\end{theorem}
Theorem \ref{Schur} could be immediately applied to $G\in SU(2)$.

Let us proceed with the computation case-wise. The $\widehat{A_n}$
series has 1 generator with 1 relation, thus $r=n=1$ and
$M(\widehat{A_n})$ is trivial. Now for the $\widehat{D_{2n}}$ series, we
note briefly that the usual presentation is $\widehat{D_{2n}} :=
\gen{a,b|a^{2n}=\II, b^2=a^n, bab^{-1}=a^{-1}}$ as in \cite{ZD};
however, we can see easily that the last two relations imply the
first, or explicitly: $a^{-n} := (bab^{-1})^n=b a^n b^{-1} = a^n$,
(q.v. \cite{Karp2} Example 3.1, Chap. 11), whence making $r=n=2$,
i.e., 2 generators and 2 relations, and further making
$M(\widehat{D_{2n}})$ trivial. Thus too are the cases of the 3
exceptional groups, each having 2 generators with 2 relations. In
summary then we have the following corollary of Theorem \ref{Schur},
the well-known \cite{AspinPles} result that
\begin{corollary}
All discrete finite subgroups of $SU(2)$ have second
cohomology $H^2(G,\IC^*) = \II$, and hence afford no non-trivial
discrete torsion.
\end{corollary}

It is intriguing that the above result can actually be hinted from
physical considerations without recourse to heavy mathematical
machinery. The orbifold theory for $G \subset SU(2)$
preserves an ${\cal N}=2$ supersymmetry on the world-volume of the
D3-Brane probe.
Inclusion of discrete torsion would deform the coefficients
of the superpotential. However, ${\cal N}=2$ supersymmetry is highly
restrictive and in general does not permit the existence of such
deformations. This is in perfect harmony with the triviality of the
Schur Multiplier of $G \subset SU(2)$ as presented in the above Corollary.

To address more complicated groups we need a methodology to compute
the Schur Multiplier, and we have many to our aid, for after all the
computation of $M(G)$ is a vast subject entirely by itself. We quote
one such method below, a result originally due to Schur:
\begin{theorem}
{\rm (\cite{Karp3} p54)} Let $G = F/R$ be the defining finite
presentation of $G$ with $F$ the free group of rank $n$ and $R$ is
(the normal closure of) the set of relations. Suppose $R/[F,R]$ has
the presentation $\gen{x_1,\ldots,x_m;y_1,\ldots,y_n}$ with all $x_i$ of
finite order, then
\[
M(G) \cong \gen{x_1,\ldots,x_n}.
\]
\end{theorem}

Two more theorems of great usage are the following:
\begin{theorem}{\rm (\cite{Karp3} p17)}
\label{exponent}
Let the exponent\footnote{i.e., the lowest common multiple of the orders
of the elements.} of $M(G)$ be $\exp(M(G))$, then
\[
\exp(M(G))^2 \mbox{ divides } |G|.
\]
\end{theorem}
And for direct products, another fact due to Schur,
\begin{theorem}{\rm (\cite{Karp3} p37)} \label{directprod}
\[
M(G_1 \times G_2) \cong M(G_1) \times M(G_2) \times (G_1 \otimes G_2),
\]
where $G_1 \otimes G_2$ is defined to be $\Hom_{\IZ}(G_1/G_1',G_2 /G_2')$.
\end{theorem}

With the above and a myriad of useful results (such as the Schur
Multiplier for semi-direct products), and especially with the aid of the
Computer Algebra package {\sf GAP} \cite{GAP} 
using the algorithm developed for the $p$-Sylow
subgroups of Schur Multiplier \cite{Holt}, 
we have engaged in the formidable task of giving
the explicit Schur Multiplier of the list of groups of our interest.

Most of the details of the computation we shall
leave to the appendix, to
give the reader a flavour of the calculation but not distracting him
or her from the main course of our writing.
Without much further ado then, we now proceed with the list of Schur
Multipliers for the
discrete subgroups of $SU(n)$ for $n=3,4$, i.e., the ${\cal N}=1,0$
orbifold theories.
\subsection{The Schur Multiplier of the Discrete Subgroups of $SU(3)$}
The classification of the discrete finite groups of $SU(3)$ is
well-known (see e.g. \cite{Fairbairn,HanHe,Muto} 
for a discussion thereof in the
context of string theory). It was pointed out in \cite{ZD} that the usual
classification of these groups does not include the so-called {\em
intransitive} groups (see \cite{SU4} for definitions), which are perhaps
of less mathematical interest. Of course from
a physical stand-point, they all give well-defined orbifolds. More
specifically \cite{ZD}, all the ordinary polyhedral subgroups of $SO(3)$, namely
the ordinary dihedral group $D_{2n}$ and the ordinary $E_6 \cong 
A_4 \cong \Delta(3 \times 2^2), E_7 \cong S_4 \cong \Delta(6 \times
2^2), E_8 \cong \Sigma_{60}$, due to
the embedding $SO(3) \hookrightarrow SU(3)$, are obviously
(intransitive) subgroups thereof and thus we
shall include these as well in what follows. We discuss some aspects
of the intransitives in Appendix B and are grateful to D. Berenstein
for pointing out some subtleties involved \cite{Berenstein}.
We insert one more cautionary note. The $\Delta(6n^2)$ series does not
actually include the cases for $n$ odd \cite{Muto}; therefore $n$
shall be restricted to be even.

Here then are the Schur Multipliers of the $SU(3)$ discrete subgroups.
\beq
\hspace{-0.7in}
\label{SU3}
\ba{|c|c|c|c|}
\hline
	& $G$ & $Order$ &  \mbox{Schur Multiplier } M(G) \\
\hline
\mbox{Intransitives}	& \IZ_n \times \IZ_m	& n \times m & \IZ_{\gcd(n,m)} \\ \hline
			& <\IZ_n, \widehat{D_{2m}}> & 
        \left\{\ba{lc}  n \times 4m & n~\odd \\ \frac{n}{2} \times 4m
        & n~\even \ea \right. &
	\left\{\ba{lc}  \II &  n~{\rm mod}~4 \neq 1 \\ 
        \IZ_2 &  n~{\rm mod}~4 = 0, m~\odd \\ 
        \IZ_2 \times \IZ_2 &  n~{\rm mod}~4 = 0, m~\even 
        \ea \right. \\ \hline

			& <\IZ_n, \widehat{E_6}> & 
        \left\{\ba{lc}  n \times 24 & n~\odd \\ \frac{n}{2} \times 24
        & n~\even \ea \right.  &
	\IZ_{\gcd(n,3)} 
	 \\ \hline

			& <\IZ_n, \widehat{E_7}>  &  
        \left\{\ba{lc}  n \times 48 & n~\odd \\ \frac{n}{2} \times 48
        & n~\even \ea \right.  &
	\left\{\ba{lc} \II & n~{\rm mod}~4 \neq 0 \\ \IZ_2 & n~{\rm mod}~4 = 0         \ea \right. \\ \hline

			& <\IZ_n, \widehat{E_8}>  &  
        \left\{\ba{lc}  n \times 120 & n~\odd \\ \frac{n}{2} \times 120
        & n~\even \ea \right.    &\II \\ \hline
			& \mbox{Ordinary Dihedral } D_{2n} & 2n &
	\IZ_{\gcd(n,2)}
	\\ \hline

			& <\IZ_n, D_{2m}> & 
        \left\{\ba{lc}  n \times 2m & m~\odd \\  
                        n \times 2m & m~\even, n~\odd \\
                        \frac{n}{2} \times 2m & m~\even, n~\even \\ \ea \right.  &
	\left\{\ba{lc}\ 
			\IZ_{\gcd(n,2)} & m~\odd \\ 
			\IZ_2  & m~\even, n~{\rm mod}~4 = 1,2,3 \\
			\IZ_2  &  m~{\rm mod}~4 \neq 0,  n~{\rm mod}~4 = 0 \\
			\IZ_2 \times \IZ_2  & m~{\rm mod}~4 = 0,  n~{\rm mod}~4 = 0 \\
	\ea\right. \\ \hline
\mbox{Transitives} 	& \Delta_{3n^2} & 3n^2 &
		\left\{\ba{cc}\IZ_n \times \IZ_3, &\gcd(n,3)\neq 1 \\
		\IZ_n, & \gcd(n,3)=1 \ea \right. \\ \hline
		 	& \Delta_{6n^2}~(n~\even) & 6n^2 & \IZ_2 \\ \hline
	& \Sigma_{60}\cong A_5 & 60 & \IZ_2 \\ \hline
	& \Sigma_{168} & 168 & \IZ_2 \\ \hline
	& \Sigma_{108} & 36 \times 3 & \II \\ \hline
	& \Sigma_{216} & 72 \times 3 & \II \\ \hline
	& \Sigma_{648} & 216 \times 3 & \II \\ \hline
	& \Sigma_{1080}& 360 \times 3 & \IZ_2 \\ \hline
\ea
\eeq
\newpage
Some immediate comments are at hand.
The question of whether any discrete subgroup of $SU(3)$ admits
non-cyclic discrete torsion was posed in \cite{AspinPles}. From our
results in table \eref{SU3}, we have shown by explicit construction
that the answer is in the affirmative: not only the various
intransitives give rise to product cyclic Schur Multipliers, so too
does the transitive $\Delta(3n^2)$ series for $n$ a multiple of 3.

In Appendix A we shall present the calculation for
$M(\Delta_{3n^2})$ and $M(\Delta_{6n^2})$ for illustrative
purposes. Furthermore, as an example of non-Abelian orbifolds with
discrete torsion, we shall investigate the series of the ordinary
dihedral group in detail with applications to physics in mind. For
now, for the reader's edification or amusement, let us continue
with the $SU(4)$ subgroups. 
\subsection{The Schur Multiplier of the Discrete Subgroups of $SU(4)$}
The discrete finite subgroups of $SL(4,\IC)$, which give rise
to non-supersymmetric orbifold theories, are presented in modern
notation in \cite{SU4}. Using the notation therein, and recalling that
the group names in $SU(4) \subset SL(4,\IC)$ were accompanied with a
star ({\it cit. ibid.}), let us tabulate the Schur Multiplier of the
exceptional cases of these particulars (cases XXIX$*$ and XXX$*$ were
computed by Prof. H. Pahlings to whom we are grateful):
\beq
\label{SU4}
\ba{|c|c|c|}
\hline
	$G$ & $Order$ & \mbox{Schur Multiplier } M(G) \\
\hline
	 $I$* & 60\times 4 &  \II \\ \hline
	 $II$* \cong \Sigma_{60} & 60 & \IZ_2 \\ \hline
	 $III$* & 360 \times 4 & \IZ_3  \\ \hline
	 $IV$* & \frac12 7!\times 2  & \IZ_3 \\ \hline
	 $VI$*   & 2^6 3^4 5\times 2 &  \II \\ \hline
	 $VII$*   & 120\times 4 & \IZ_2 \\ \hline
	 $VIII$* & 120\times 4	& \IZ_2	\\ \hline	
	 $IX$*   & 720\times 4	& \IZ_2	\\ \hline
	 $X$*    & 144 \times 2 & \IZ_2 \times \IZ_3 \\ \hline
	 $XI$*   & 288 \times 2 & \IZ_2 \times \IZ_3 \\ \hline
	 $XII$*  & 288 \times 2 & \IZ_2 \\ \hline
	 $XIII$* & 720 \times 2 & \IZ_2 \\ \hline
	 $XIV$*  & 576 \times 2 & \IZ_2 \times \IZ_2 \\ \hline
	 $XV$*   & 1440 \times 2 & \IZ_2 \\ \hline
\ea
\quad
\ba{|c|c|c|}
\hline
	$G$ & $Order$ & \mbox{Schur Multiplier } M(G) \\
\hline
	 $XVI$*  & 3600 \times 2 & \IZ_2 \\ \hline
	 $XVII$* &576\times 4 & \IZ_2	\\ \hline
	 $XVIII$*&576\times 4 & \IZ_2 \times \IZ_3 \\ \hline
	 $XIX$*  &288\times 4 & \II	\\ \hline
	 $XX$*   &7200\times 4 & \II	\\ \hline
	 $XXI$*  &1152\times 4 & \IZ_2 \times \IZ_2 \\ \hline
	 $XXII$* & 5\times 16 \times 4   & \IZ_2 \\ \hline
	 $XXIII$*& 10\times 16 \times 4  & \IZ_2 \times \IZ_2 \\ \hline
	 $XXIV$* & 20\times 16 \times 4  & \IZ_2 \\ \hline
	 $XXV$*  & 60\times 16 \times 4  & \IZ_2 \\ \hline
	 $XXVI$* & 60\times 16 \times 4  & \IZ_2 \times \IZ_4 \\ \hline
	 $XXVII$*& 120\times 16 \times 4 & \IZ_2 \times \IZ_2 \\ \hline
	 $XXVIII$*& 120\times 16 \times 4& \IZ_2 \\ \hline
	 $XXIX$* & 360\times 16 \times 4 & \IZ_2 \times \IZ_3\\ \hline
	 $XXX$*  & 720\times 16 \times 4 & \IZ_2 \\ \hline
\ea
\eeq
\newpage
\section{$D_{2n}$ Orbifolds: Discrete Torsion for a
non-Abelian Example}
As advertised earlier at the end of subsection 3.2, 
we now investigate in depth the discrete
torsion for a non-Abelian orbifold.
The ordinary dihedral group $D_{2n} \cong \IZ_n \rtimes \IZ_2$ of order
$2n$, has the presentation
\[
D_{2n}=\gen{a,b|a^n=1, b^2=1, bab^{-1}=a^{-1}}.
\]
As tabulated in \eref{SU3}, the Schur Multiplier is $M(D_{2n})=\II$ 
for $n$ odd and $\IZ_2$ for $n$ even \cite{Karp}.
Therefore the $n$ odd cases are no different from the ordinary linear
representations as studied in \cite{ZD} since they have trivial Schur
Multiplier and hence trivial discrete torsion. On the other hand, for
the $n$ even case, we will demonstrate the following result:
\begin{proposition}
The binary dihedral group $\widehat{D_{2n}}$ of the
$D$-series of the discrete subgroups of $SU(2)$ (otherwise
called the generalised quaternion group) 
is the covering group of $D_{2n}$ when $n$ is even. 
\end{proposition}
Proof: The definition of the binary dihedral group $\widehat{D_{2n}}$, of
order $4n$, is
\[
\widehat{D_{2n}}=\gen{a,b|a^{2n}=1, b^2=a^n, bab^{-1}=a^{-1}},
\]
as we saw in subsection 3.1.
Let us check against the conditions of Theorem \ref{cover}.
It is a famous result that $\widehat{D_{2n}}$ is the double cover of
$D_{2n}$ and whence an $\IZ_2$ central extension.
First we can see that $A=Z(\widehat{D_{2n}})=\{1,a^n\}\cong \IZ_2$ and
condition (ii) is satisfied.
Second we find that the commutators are $[a^x,a^y] :=
(a^x)^{-1}(a^y)^{-1} a^x a^y=1$, $[a^xb, a^yb]=a^{2(x-y)}$ 
and $[a^x b, a^y]=a^{2y}$. From these we see that the derived group
$[\widehat{D_{2n}},\widehat{D_{2n}}]$ is generated by $a^2$ and is thus
equal to $\IZ_n$ (since $a$ is of order $2n$). An important point is that
only when $n$ is even does $A$ belong to $Z(\widehat{D_{2n}}) \cap
[\widehat{D_{2n}},\widehat{D_{2n}}]$. This result is consistent with the fact that
for odd $n$, $D_{2n}$ has trivial Schur Multiplier. Finally of course, 
$|A| = |H^2(G,\IF^*)| = 2$. Thus conditions (i)
and (iii) are also satisfied.
We therefore conclude that for even $n$, 
$\widehat{D_{2n}}$ is the covering group of $D_{2n}$.
\subsection{The Irreducible Representations}
With the above Proposition, we know by the very definition of the
covering group, that the projective representation of $D_{2n}$ should be
encoded in the linear representation of $\widehat{D_{2n}}$, which is a
standard result that we can recall from \cite{ZD}.
The latter has four 1-dimensional and $n-1$ 2-dimensional irreps.
The matrix representations of these 2-dimensionals for the generic
elements $a^p, b a^p$ ($p=0,...,2n-1$) are given below:
\beq 
\label{binarydihedral_2}
a^{p} = \mat{\ba{cc}  \omega_{2n}^{lp} & 0 \\
				0 & \omega_{2n}^{-lp}
		\ea}
\qquad
b a^{p} = \mat{\ba{cc}  0 & i^l\omega_{2n}^{-lp} \\
				i^l \omega_{2n}^{lp} & 0 
		\ea},
\eeq
with $l=1,...,n-1$; these are denoted as $\chi^l_2$. 
On the other hand, the four 1-dimensionals are
\beq
{\small
\label{binarydihedral_1}
\ba{c|c}
n = \even
&
n = \odd \\ \hline
\ba{c|cccc}
     &	a^{\even} & a(a^{\odd}) & b(b a^{\even}) &
	b a (b a^{\odd}) \\
\chi^{1}_1 & 1 & 1 & 1 & 1 \\
\chi^{2}_1 & 1 & -1 & 1 & -1 \\
\chi^{3}_1 & 1 & 1 & -1 & -1 \\
\chi^{4}_1 & 1 & -1 & -1 & 1 
\ea
&
\ba{cccc}
	a^{\even} & a(a^{\odd}) & b(b a^{\even}) &
	b a (b a^{\odd}) \\
1 & 1 & 1 & 1 \\
1 & -1 & \omega_4 & -\omega_4 \\
1 & 1 & -1 & -1 \\
1 & -1 & - \omega_4 &  \omega_4
\ea
\ea}
\eeq

We can subsequently obtain all irreducible projective
representations of $D_{2n}$ from the above (henceforth $n$ will be 
even).
Recalling that $\widehat{D_{2n}}/\{1,a^n\}\cong D_{2n}$ from property (ii) of
Theorem \ref{cover}, we can choose one element of each of the
transversals of $\widehat{D_{2n}}$ with respect to the $\IZ_2$ to be
mapped to $D_{2n}$.
For convenience we choose $b^{x} a^{y}$ with $x=0,1$ and
$y=0,1,...,n-1$, a total of $4n/2=2n$ elements. Thus we are
effectively expressing $D_{2n}$ in terms of $\widehat{D_{2n}}$ elements.

For the matrix representation of $a^n \in \widehat{D_{2n}}$, there are two cases. 
In the first, we have $a^n=1\times I_{d\times d}$ where $d$ is the dimension
of the representation. This case includes all four 1-dimensional representations
and $(n/2-1)$ 2-dimensional representations in \eref{binarydihedral_2}
for $l=2,4,...,n-2$. Because $a^n$ has the same matrix form as $\II$, we see that
the elements $b^{x} a^{y}$ and $b^{x} a^{y+n}$ also have the same
matrix form. Consequently, when we map them to $D_{2n}$, they automatically give the
irreducible linear representations of $D_{2n}$.

In the other case, we have $a^n=-1\times I_{d\times d}$ and this happens
when $l=1,3,...,n-1$. It is precisely these cases\footnote{Sometimes
	also called {\bf negative representations} in such cases.}
which give the {\em irreducible projective representations} of
$D_{2n}$. Now, because $a^n$ has a different matrix form
from $\II$, the matrices for $b^{x} a^{y}$ and $b^{x} a^{y+n}$ differ.
Therefore, when we map $\widehat{D_{2n}}$ to $D_{2n}$, there
is an ambiguity as to which of the matrix forms,
$b^{x} a^{y}$ or $b^{x} a^{y+n}$,
to choose as those of $D_{2n}$.

This ambiguity is exactly a feature of projective representations.
Preserving the notations of Theorem \ref{cover}, we let $G^* =
\bigcup\limits_{g_i \in G} A g_i$ be the decomposition into
transversals of $G$ for the normal subgroup $A$. Then choosing 
one element in every transversal, say $A_q g_i$ for some fixed $q$, we
have the ordinary (linear) representation thereof being precisely the
projective representation of $g_i$. Of course different choices of
$A_q$ give different but projectively equivalent (projective)
representations of $G$.

By this above method, we can construct all irreducible projective
representations of $D_{2n}$ from \eref{binarydihedral_2}.
We can verify this by matching dimensions:
we end up with $n/2$ 2-dimensional
representations inherited from $\widehat{D_{2n}}$ and
$2^2 \times (n/2)=2n$, which of course is the order of $D_{2n}$ as it
should.
\subsection{The Quiver Diagram and the Matter Content}
The projection for the matter content $\Phi$ is well-known (see
e.g., \cite{LNV,HanHe}):
\begin{equation}
\label{projection}
\gamma^{-1}(g) \Phi \gamma(g)= r(g) \Phi,
\end{equation}
for $g \in G$ and $r, \gamma$ appropriate (projective) representations.
The case of $D_{2n}$ without torsion was discussed as a new class of
non-chiral ${\cal N}=1$ theories in \cite{ZD}. We recall that
for the group $D_{2n}$ we choose the generators (with action on
$\IC^3$) as
\begin{equation} 
\label{dihedral_2}
a = \left(  \begin{array}{ccc} 1 & 0 & 0 \\0 &  \omega_{n} & 0 \\
				0 &	0 & \omega_{n}^{-1}
			\end{array}
		\right)
\qquad
b  = \left(  \begin{array}{ccc}  -1 & 0 & 0\\ 0 & 0 & -1 \\
				0 & -1 & 0 
			\end{array}
		\right).
\end{equation}
Now we can use these explicit forms to work out the matter content
(the quiver diagram) and superpotential. For the regular
representation, we
choose $\gamma(g)$ as block-diagonal in which every 2-dimensional irreducible 
representation repeats twice with labels $l=1,1,3,3,..
,n-1,n-1$ (as we have shown in the previous section that the even
labels correspond to the linear representation of $D_{2n}$).
With this $\gamma(g)$, we calculate the matter content below.

For simplicity, in the actual calculation we would not use 
\eref{projection} but rather the standard method given by Lawrence,
Nekrasov and Vafa \cite{LNV}, generalised appropriately
to the projective case by
\cite{AspinPles}. We can do so because we are armed with
Definition \ref{defcover} and results from the previous subsection,
and directly use the linear representation of the covering group:
we lift the action of $D_{2n}$ into the action of its covering group
$\widehat{D_{2n}}$. It is easy to see that we get the 
same matter content either by using the projective representations of
the former or the linear representations of the latter.

From the point of view of the covering group, the representation $r(g)$ in 
\eref{projection} is given by
\beq\label{decomp}
{\bf 3} \longrightarrow \chi^3_1+\chi^2_2
\eeq
and the representation $\gamma(g)$ is given by
$\gamma \longrightarrow \sum\limits_{l=0}^{n/2-1} 2 \chi^{2l+1}_2$.
We remind ourselves that the ${\bf 3}$ must in fact be a {\em linear}
representation of $D_{2n}$ while $\gamma(g)$ is the one that has to be {\em
projective} when we include discrete torsion \cite{Doug}.

For the purpose of tensor decompositions we recall the result for the binary
dihedral group \cite{ZD}:\\
\begin{equation}
\begin{array}{|l|l|}
\hline
{\bf 1} \otimes {\bf 1}'
&
\begin{array}{c|c}
	n = \even	& n = \odd \\
	\begin{array}{ccc}
	\chi_1^2\chi_1^2=\chi_1^1  & \chi_1^3\chi_1^3=\chi_1^1 &
	\chi_1^4\chi_1^4=\chi_1^1  \\
	\chi_1^2 \chi_1^3=\chi_1^4 & \chi_1^2\chi_1^4=\chi_1^3 &
	\chi_1^3\chi_1^4=\chi_1^2
	\end{array}
	&
	\begin{array}{ccc}
	\chi_1^2\chi_1^2=\chi_1^3  & \chi_1^3\chi_1^3=\chi_1^1 &
	\chi_1^4\chi_1^4=\chi_1^3  \\
	\chi_1^2 \chi_1^3=\chi_1^4 & \chi_1^2\chi_1^4=\chi_1^1 &
	\chi_1^3\chi_1^4=\chi_1^2
	\end{array}
\end{array}
\\ \hline
{\bf 1} \otimes {\bf 2}
&
\chi_1^{h} \chi_2^l = \left\{ \begin{array}{l}
\chi_2^l~~~~h=1,3  \\
\chi_2^{n-l}~~~~h=2,4 
\end{array}
\right.
\\ \hline
{\bf 2} \otimes {\bf 2'}
&
\chi_2^{l_1} \chi_2^{l_2}=\chi_2^{(l_1+l_2)}+\chi_2^{(l_1-l_2)}
{\rm ~where~}
\begin{array}{l}
	\chi_2^{(l_1+l_2)}= \left\{ \begin{array}{l}
	\chi_2^{(l_1+l_2)}~~~~{\rm if}~~~l_1+l_2<n,  \\
	\chi_2^{2n-(l_1+l_2)}~~~~{\rm if}~~~l_1+l_2>n, \\
	\chi_1^2+\chi_1^4~~~~{\rm if}~~~l_1+l_2=n.
	\end{array}
	\right.
	\\
	\chi_2^{(l_1-l_2)}= \left\{ \begin{array}{l}
	\chi_2^{(l_1-l_2)}~~~~{\rm if}~~~l_1>l_2,  \\
	\chi_2^{(l_2-l_1)}~~~~{\rm if}~~~l_1<l_2, \\
	\chi_1^1+\chi_1^3~~~~{\rm if}~~~l_1=l_2.
	\end{array}
	\right.
\end{array}
\\
\hline
\end{array}
\end{equation}

From these relations we immediately obtain the matter content.
Firstly, there are $n/2$ $U(2)$ gauge groups ($n/2$ nodes in the quiver). 
Secondly, because $\chi^3_1 \chi_2^l=\chi_2^l$ we have one adjoint 
scalar for every gauge group. Thirdly, since $\chi_2^{2} \chi_2^{2l+1}=
\chi_2^{2l-1}+\chi_2^{2l+3}$ (where for $l=0$, 
$\chi_2^{2l-1}$ is understood to
be $\chi_2^{1}$ and for $l=n/2-1$, $\chi_2^{2l+3}$ is understood to be
$\chi_2^{n-1}$), we have two bi-fundamental chiral supermultiplets. 
We summarise these results in \fref{fig:dihedral}.
\EPSFIGURE[ht]{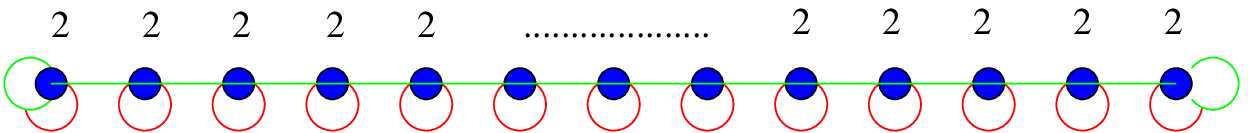,width=6.2in}
{
The quiver diagram of the ordinary dihedral group $D_{2n}$
with non-trivial projective representation. In this case of
discrete torsion being turned on, we have a product of $n/2$ $U(2)$
gauge groups (nodes).
The line connecting two nodes without arrows means
that there is one chiral multiplet in each direction. Therefore we
have a non-chiral theory.
\label{fig:dihedral}
}

We want to emphasize that by lifting to the covering group, in
general we not only find the matter content (quiver diagram) as we
have done above, but also the superpotential as well. The formula is
given in (2.7) of \cite{LNV}, which could be applied here without
any modification (of course, one can use the matrix form of the group
elements to obtain the superpotential directly as
done in \cite{Doug,DougFiol,DM,BL,BJL,Ray,Klein}, 
but (2.7), expressed in terms of the Clebsh-Gordan coefficients, is more 
convenient).

Knowing the above quiver (cf. \fref{fig:dihedral}) of the ordinary dihedral
group $D_{2n}$ {\em with} discrete torsion, we wish to question
ourselves as to the relationships between this quiver and that of
its covering group, the
binary dihedral group $\widehat{D_{2n}}$ {\em without} discrete
torsion (as well as that of $D_{2n}$ without discrete 
torsion).
The usual quiver of $\widehat{D_{2n}}$ is well-known \cite{JM,HanHe};
we give an example for $n=4$ in part (a) of \fref{fig:comparing}.
The quiver is obtained 
by choosing the decomposition of ${\bf 3}\longrightarrow 
\chi_1^1+\chi_2^1$ (as opposed to \eref{decomp} because this is the
linear representation of $\widehat{D_{2n}}$); also
$\gamma(g)$ is in the regular representation of dimension $4n$.
A total of $(n-1)+4=n+3$ nodes results.
We recall that when getting the quiver of $D_{2n}$
with discrete torsion in the above, we chose the decomposition of 
${\bf 3} \longrightarrow\chi_1^3+\chi_2^2$ in \eref{decomp} which
provided a linear representation of $D_{2n}$.
Had we made this same choice for $\widehat{D_{2n}}$, our familiar quiver 
of $\widehat{D_{2n}}$ would have split into two parts: 
one being precisely the quiver of  $D_{2n}$ 
without discrete torsion as discussed in \cite{ZD} and the other, 
that of $D_{2n}$ with discrete
torsion as presented in \fref{fig:dihedral}. 
These are given respectively in
parts (b) and (c) of \fref{fig:comparing}.

From this discussion, we see that in some sense discrete torsion is
connected with different choices of decomposition in the usual orbifold
projection. We want to emphasize that the example of $D_{2n}$ is very
special because its covering group $\widehat{D}_{2n}$ belongs to $SU(2)$.
In general, the covering group does not even belong to $SU(3)$ and the meaning
of the usual orbifold projection of the covering group in string theory is
vague.
\EPSFIGURE[ht]{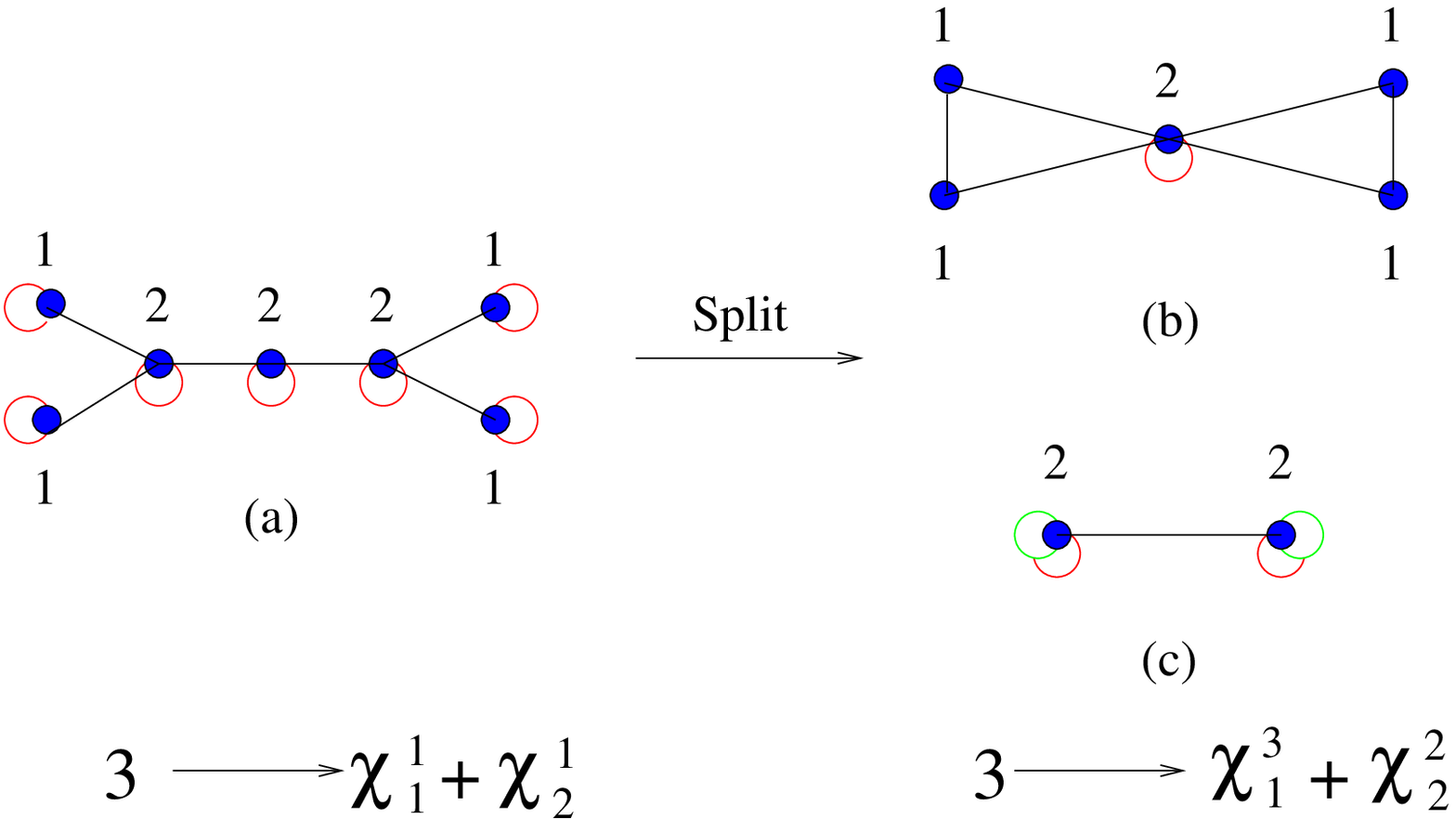,width=4.5in}
{
(a) The quiver diagram of the binary dihedral group 
$\widehat{D}_4$ {\em without}
discrete torsion; (b) the quiver of the ordinary dihedral group 
$D_4$ {\em without}
discrete torsion; (c) the quiver of the ordinary dihedral group 
$D_4$ {\em with} discrete torsion.  
\label{fig:comparing}
}
\section{Conclusions and Prospects}
Let us pause here awhile for reflection. A key purpose of this writing
is to initiate the investigation of discrete torsion for the generic
D-brane orbifold theories. Inspired by this goal, we have shown that
computing the Schur Multiplier $M(G)$ for the finite group $G$
serves as a beacon in our quest.

In particular, using the fact that $M(G)$ is an indicator of when we
can turn on a non-trivial 
NS-NS background in the orbifold geometry and when we cannot: only when
$M(G)$, as an Abelian group is not trivially $\II$ can the former be
executed. As a guide for future investigations, we have computed
$M(G)$ for the discrete subgroups $G$ in $SU(n)$ with $n=2,3,4$, which
amounts to a {\em classification of which D-brane orbifolds afford
non-trivial discrete torsion}.

As an explicit example, in supplementing the present lack of studies
of non-Abelian orbifolds with discrete torsion in the current literature,
we have pursued in detail the ${\cal N}=1$ gauge
theory living on the D3-Brane probe on the orbifold singularity
$\IC^3/D_{2n}$, corresponding to the ordinary dihedral group of order $2n$
as a subgroup of $SU(3)$. As the group has Schur Multiplier $\IZ_2$
for even $n$, we have turned on the discrete torsion and arrived at an
interesting class of non-chiral theories.

The prospects are as manifold as the interests are diverse and much
work remains to be done. An immediate task is
to examine the gauge theory living on the world-volume of D-brane probes
when we turn on the discrete torsion of a given
orbifold wherever allowed by our classification.
This investigation is currently in progress.

Our results of the Schur Multipliers could also be interesting to the
study of K-theory in connexion to string theory.
Recent works \cite{Gomis,AspinPles,Kapustin} have noticed an
intimate relation between twisted K-theory and discrete torsion. 
More specifically, the Schur Multiplier of an orbifold group may
in fact supply information about the torsion subgroup of the cohomology
group of space-time in the light of a generalised McKay Correspondence
\cite{AspinPles,HeSong}.

It is also tempting to further study the non-commutative moduli
space of non-Abelian orbifolds in the spirit of \cite{DougFiol,BL,BJL}
which treated Abelian cases at great length. How the framework
developed therein extends to the non-Abelian groups should be interesting.
Works on discrete torsion in relation to permutation orbifolds and
symmetric products \cite{Sym} have also been initiated, we hope that our
methodologies could be helpful thereto.

Finally, there is another direction of future study. 
The boundary state formalism was used in
\cite{Gab} where it was suggested that the ties between close and
open string sectors maybe softened with regard to discrete torsion.
It is thus natural to ask if such ambiguities may exist also
for non-Abelian orbifolds.

All these open issues, of concern to the physicist and the
mathematician alike, present themselves to the intrigue of the
reader.
\section*{Acknowledgements}
{\it Ad Catharinae Sanctae Alexandriae et Ad Majorem Dei Gloriam...\\}
We would like to acknowledge Professors H. Pahlings and J. Neubueser
of the Lehrstuhl D f\"ur Mathematik, Germany, for their tireless
support. Moreover, we thank Prof. M. Artin of the Dept. of
Mathematics, MIT, for his helpful insights and Prof. J. Humphreys of
the University of Liverpool, UK, for comments. In addition we are
indebted to P. Aspinwall, G. Bertoldi, B. Fiol and N. Moeller
for discussions as well as
B. Pearre and M. Warden of the Dept. of Brain and Cognitive Sciences,
MIT, for lending a hand at computer resources. We are also
grateful to M. Warden for suggestions to the last draft of the
manuscript.
Finally we are particularly obliged to D. Berenstein for careful
examination of the first versions of the paper and pointing out
errors therein.
\section{Appendix A: Some Explicit Computations for $M(G)$}
\subsection{Preliminary Definitions}
We begin with a few rudimentary definitions \cite{Karp}.
Let $H$ be a subgroup of $G$ and let $g \in G$. For any cocycle
$\alpha \in  Z^2(G,\IC^*)$ we define an induced action
$g \cdot \alpha \in Z^2(g H g^{-1}, \IC^*)$ thereon as
$g \cdot \alpha(x,y) = \alpha(g^{-1} x g, g^{-1} y g),~~\forall~~x,y 
\in gHg^{-1}$.
Now, it can be proved that the mapping
\[
c_{g} : M(H) \rightarrow M(gHg^{-1}), \;\;\; c_{g}(\alpha) := g \cdot \alpha
\]
is a homomorphism, which we call {\bf cocycle conjugation} by $g$.

On the other hand we have an obvious concept of restriction:
for $S \subseteq L$ subgroups of $G$, we denote by  Res$_{L,S}$ the
restriction map $M(L) \rightarrow M(S)$. Thereafter we define
stability as:
\begin{definition}
Let $H$ and $K$ be arbitrary subgroups of $G$. An element
$\alpha \in M(H)$ is said to be {\bf K-stable} if
\[
{\rm Res}_{H,gHg^{-1}\cap H}(\alpha) = 
{\rm Res}_{gHg^{-1},gHg^{-1}\cap H}(c_{g}(\alpha)) ~~\forall~~ g \in K.
\] 
\end{definition}
The set of all K-stable elements of $M(H)$ will be denoted
by $M(H)^{K}$ and it forms a subgroup of $M(H)$ known
as the K-stable subgroup of $M(H)$.

When $K \subseteq N_{G}(H)$ all the above concepts\footnote{$N_{G}(H)$
	is the normalizer of $H$ in $G$, i.e., the set of all elements
	$g \in G$ such that $g H g^{-1} = H$. When $H$ is a normal
	subgroup of $G$ we obviously have $N_{G}(H)=G$.}
	coalesce and we have the following important lemma:
\begin{lemma}
\label{lemma_stab}{\rm (\cite{Karp} p299)}
If $H$ and $K$ are subgroups of $G$ such that $K \subseteq N_{G}(H)$,
then $M(H)^{K}$ is the K-stable subgroup of $M(H)$ with respect to
the action of $K$ on $M(H)$ induced by the action of $K$ on $H$
by conjugation. In other words,
\[
M(H)^{K}=\{\alpha \in M(H), \;\;   \alpha(x,y)=c_{g}(\alpha)(x,y)
~~\forall~~ g \in K, ~~\forall~~ x,y \in H\}.
\]
\end{lemma}

Finally let us present a useful class of subgroups:
\begin{definition}
A subgroup $H$ of a group $G$ is called a {\bf Hall subgroup} of $G$
if the order of $H$ is coprime with its index in $G$,
i.e. $\gcd(|H|,|G/H|)=1$.
\end{definition}
For these subgroups we have:
\begin{theorem}{\rm (\cite{Karp} p334)}
\label{normalhall}
If $N$ is a normal Hall subgroup of $G$. Then
\[
M(G) \cong M(N)^{G/N} \times M(G/N).
\]
\end{theorem}

The above theorem is really a corollary of a more general case of
semi-direct products:
\begin{theorem}\label{sequence}{\rm (\cite{Karp3} p33)}
Let $G = N \rtimes T$ with $N \triangleleft G$, then \\
$(i) \quad M(G) \cong M(T) \times \tilde{M}(G)$; \\
$(ii) \quad \mbox{The sequence }
	1 \rightarrow H^1(T,N^*) \rightarrow \tilde{M}(G)
	{\stackrel{{\rm Res}}{\rightarrow}} M(N)^T \rightarrow H^2(T,N^*)
	\mbox{ is exact,}$\\
where $\tilde{M}(G) := \ker{\rm Res}_{G,N}$, $N^* := \Hom(N,\IC^*)$ and
$H^{i=1,2}(T,N^*)$ is the cohomology defined with respect to the
conjugation action by $T$ on $N^*$.
\end{theorem}
Part (ii) of this theorem actually follows from the
Lyndon-Hochschild-Serre spectral sequence into which we shall not delve.

One clarification is needed at hand. Let us define the first
$A$-valued cohomology group for $G$, which we shall utilise later in our
calculations. Here the 1-cocycles are the set of functions $Z^1(G,A) :=
\{f:G\rightarrow A|f(xy) = (x\cdot f(y))f(x)\quad\forall x,y \in G\}$,
where $A$ is being acted upon ($x \cdot A \rightarrow A$ for $x \in G$) 
by $G$ as a $\IZ G$-module. These are known as {\em crossed
homomorphisms}. On the other hand, the 1-coboundaries are what is
known as the principal crossed homomorphisms, $B^1(G,A) := \{f_{a \in
A}(x) = (x \cdot a) a^{-1} \}$ from which we define $H^1(G,A) :=
Z^1(G,A) / B^1(G,A)$.

Alas, {\it caveat emptor}, we have defined in subsection 2.2,
$H^2(G,A)$. There, the action of $G$ on $A$ (as in the case of the Schur
Multiplier) is taken to be trivial, we must be careful, in
the ensuing, to compute with respect to non-trivial actions such as
conjugation. In our case the conjugation action of $t \in T$ on
$\chi \in \Hom(N,\IC^*)$ is given by $\chi(t n t^{-1})$ for $n\in N$.
\subsection{The Schur Multiplier for $\Delta_{3n^2}$}
\subsubsection{Case I: $\gcd(n,3)=1$}
Thus equipped, we can now use theorem \ref{normalhall} at our ease to
compute the Schur multipliers the first case of
the finite groups $\Delta_{3 n^2}$. Recall that
$\IZ_{n} \times \IZ_{n} \triangleleft \Delta(3n^2)$ or explicitly
\[
\Delta_{3 n^2} \cong (\IZ_{n} \times \IZ_{n}) \rtimes \IZ_{3}.
\]
Our crucial observation is that when $\gcd(n,3)=1$, $\IZ_{n} \times
\IZ_{n}$ is in fact a normal Hall subgroup of $\Delta_{3 n^2}$ with
quotient group $\IZ_{3}$. Whence Theorem \ref{normalhall} can be
immediately applied to this case when  $n$ is coprime to 3:
\[
M(\Delta_{3 n^2})=(M(\IZ_n\times\IZ_n))^{\IZ_3} \times M(\IZ_3) 
		= (M(\IZ_n\times\IZ_n))^{\IZ_3},
\]
by recalling that the Schur Multiplier of all cyclic groups is
trivial and that of $\IZ_n\times\IZ_n$ is $\IZ_n$ \cite{Karp}.
But, $\IZ_3 \subseteq N_{\Delta_{3 n^2}}(\IZ_{n} \times
\IZ_{n})=\Delta_{3 n^2}$, and hence by Lemma \ref{lemma_stab} it suffices
to compute the $\IZ_3$-stable subgroup of $\IZ_n$ by cocycle conjugation.

Let the quotient group $\IZ_3$ be $\gen{z|z^3=\II}$ and similarly, if
$x, y, x^n=y^n=\II$ are the generators of $\IZ_{n} \times \IZ_{n}$,
then a generic element thereof becomes $x^a y^b, a,b = 0,\ldots,n-1$.
The group conjugation by $z$ on such an element gives
\beq\label{conj}
z^{-1} x^a y^b z = x^b y^{-a-b} \qquad z x^a y^b z^{-1} = x^{-a-b} y^a.
\eeq
It is easy now to check that if $\alpha$ is
a generator of the Schur multiplier $\IZ_{n}$, we have an induced action
\[
c_{z}(\alpha)(x^a y^b,x^{a'} y^{b'}):=
\alpha(z^{-1} x^a y^b z,z^{-1} x^{a'} y^{b'} z)=
\alpha(x^b y^{-(a+b)}, x^{b'} y^{-(a'+b')})
\]
by Lemma \ref{lemma_stab}.

However, we have a well-known result \cite{Klein}:
\begin{proposition}\label{zeta}
For the group $\IZ_{n} \times \IZ_{n}$, the explicit generator of the
Schur Multiplier is given by
\[
\alpha(x^a y^b, x^{a'} y^{b'})=\omega_{n}^{a b' - a' b}.
\]
\end{proposition}
Consequently, $\alpha(x^b y^{-(a+b)}, x^{b'} y^{-(a'+b')})=
\alpha(x^a y^b,x^{a'} y^{b'})$ whereby making the $c_z$-action trivial
and causing $(M(\IZ_n \times \IZ_n)^{\IZ_3} \cong M(\IZ_n \times
\IZ_n) = \IZ_n$. From this we conclude part
I of our result: $M(\Delta_{3 n^2}) = \IZ_{n}$ for
$n$ coprime to 3.
\subsubsection{Case II: $\gcd(n,3) \neq 1$}
Here the situation is much more involved. Let us appeal to Part (ii)
of Theorem \ref{sequence}. We let $N = \IZ_n \times \IZ_n$ and $T =
\IZ_3$ as above and define $U := \Hom(\IZ_n \times \IZ_n,\IC^*))$;
the exact sequence then takes the form
\beq\label{seq2}
1 \rightarrow H^1(\IZ_3,U) \rightarrow \tilde{M}(\Delta_{3n^2}) \rightarrow \IZ_n 
 \rightarrow H^2(\IZ_3,U)
\eeq
using the fact that the stable subgroup $M(\IZ_n \times
\IZ_n)^{\IZ_3} \cong \IZ_n$ as shown above.
Some explicit calculations are now called for.

As for $U$, it is of course isomorphic to $\IZ_n \times \IZ_n$ since
for an Abelian group $A$, $\Hom(A,\IC^*) \cong A$ (\cite{Karp3}
p17). We label the elements thereof as $(p,q)(x^a y^b) := \omega_n^{a
p + b q}$, taking $x^a y^b \in \IZ_n \times \IZ_n$ to $\IC^*$.

We recall that the conjugation by $z \in \IZ_3$ on $\IZ_n \times
\IZ_n$ is \eref{conj}. Therefore, by the remark at the end of the
previous subsection, $z$ acts on $U$ as: $(z \cdot (p,q))(x^a y^b) 
:= (p,q)(z(x^a
y^b)z^{-1}) = \omega_n^{a' p + b' q}$ with $a' = -a-b$ and $b' = a$
due\footnote{Note that we must be careful to let the order of conjugation
	be the opposite of that in the cocycle conjugation.} 
to \eref{conj}, whence 
\beq\label{conjZ3}
z \cdot (p,q) = (q-p,-p),\quad\mbox{ for }(p,q) \in U. 
\eeq

Some explicit calculations are called for.
First we compute $H^1(\IZ_3,U)$. $Z^1$ is
generically composed of functions such that $f(z) = (p,q)$
(and also $f(\II) = \II$ and $f(z^2) = (z \cdot f(z)) f(z)$ by the
crossed homomorphism condition, and is subsequently equal to
$(q,p+q)$ by \eref{conjZ3}. Since no
further conditions can be imposed, $Z^1 \cong \IZ_n \times \IZ_n$.
Now $B^1$ consists of all functions of the form  $(z \cdot
(p,q))(p,q)^{-1} = (q-2p,-p-q)$, these are to be identified
with the trivial map in $Z^1$. We can re-write these elements as 
$(p':=q-2p,-p'-3p) = (\omega_n^a\omega_n^{-b})^{p'}(\omega_n^b)^{-3p}$,
and those in $Z^1$ we re-write as
$(\omega_n^a\omega_n^{-b})^{p'}(\omega_n^b)^{q'}$ as we are free to do.
Therefore if $\gcd(3,n)=1$, then 
$H^1 := Z^1/B^1$ is actually trivial because in mod $n$, $3p$ also
ranges the full $0,\cdots, n-1$, whereas if $\gcd(3,n) \ne 1$ then 
$H^1 := Z^1/B^1 \cong \IZ_3$.

The computation for $H^2(\IZ_3,U)$ is a little more
involved, but the idea is the same. First we determine $Z^2$ as
composed of $\alpha(z_1,z_2)$ constrained by the cocycle condition (with
respect to conjugation which differs from \eref{alpha} where the trivial
action was taken)
\[
\alpha(z_1,z_2) \alpha(z_1 z_2,z_3) = (z_1 \cdot
\alpha(z_2,z_3)) \alpha(z_1,z_2 z_3) \qquad z_1,z_2,z_3 \in \IZ_3.
\] 
Again we
only need to determine the following cases: $\alpha(z,z) :=
(p_1,q_1); \alpha(z^2,z^2) := (p_2,q_2);$ $\alpha(z^2,z) :=
(p_3,q_3); \alpha(z,z^2) := (p_4,q_4)$. The cocycle
constraint gives $(p_1,q_1) = (q_4,-q_3);$ $(p_2,q_2) = (-q_3-q_4,-q_4);
(p_3,q_3) = (-q_4,q_3); (p_4,q_4) = (p_4,q_4)$, giving
$Z^2 \cong \IZ_n \times \IZ_n$.
The coboundaries are given by $(\delta t)(z_1,z_2) = (z_1 \cdot t(z_2))
t(z_1) t(z_1 z_2)^{-1}$ (for any mapping $t : \IZ_3 \rightarrow \IZ_n
\times \IZ_n$ which we define to take values $t(z) = (r_1,s_1)$ and
$t(z^2) = (r_2,s_2)$)), making $(\delta t)(z,z) = (s_1-r_2,-r_1+s_1-s_2);
(\delta t)(z^2,z^2) (-s_2+r_2-r_1,r_2-s_1);
(\delta t)(z^2,z) = (-s_1+r_2,r_1-s_1+s_2);
(\delta t)(z,z^2) = (s_2-r_2+r_1,s_1-r_2)$. Now, the transformation
$r_2 = s_1 + q_4; r_1 = s_1 - s_2 - p_4 + q_4$ makes this set of
values for $B^2$ completely identical to those in $Z^2$, whence we
conclude that $B^2 \cong \IZ_n \times \IZ_n$.
In conclusion then $H^2 := Z^2 / B^2 \cong \II$.

The exact sequence \eref{seq2} then assumes the simple form of
\[
1\rightarrow \left\{\ba{cc}\IZ_3, &\gcd(n,3)\neq 1 \\
	\II, & \gcd(n,3)=1\ea \right\}\rightarrow \tilde{M}(G)
 \rightarrow \IZ_n \rightarrow 1,
\]
which means that if $n$ does not
divide 3, $\tilde{M}(G) \cong \IZ_n$, and otherwise $\tilde{M}(G) /
\IZ_3 \cong \IZ_n$. Of course, in conjunction with Part (i) of Theorem
\ref{sequence}, we immediately see that the first case makes Part I of
our discussion (when $\gcd(n,3)= 1$) a special case of our present
situation.

On the other hand, for the remaining case of $\gcd(n,3)\neq 1$, we
have $M(\Delta_{3n^2})/\IZ_3 \cong \IZ_n$, which means that
$M(\Delta_{3n^2})$, being an Abelian group, can only be $\IZ_{3n}$ or
$\IZ_n \times \IZ_3$. The exponent of the former is $3n$, while the
later (since 3 divides $n$), is $n$, but by Theorem \ref{exponent}, the
exponent squared must divide the order, which is $3n^2$, whereby
forcing the second choice.

Therefore in conclusion we have our {\it theorema egregium}:
\[
M(\Delta_{3n^2}) = \left\{\ba{cc}\IZ_n \times \IZ_3, &\gcd(n,3)\neq 1 \\
	\IZ_n, & \gcd(n,3)=1 \ea \right.
\]
as reported in Table \eref{SU3}.
\subsection{The Schur Multiplier for $\Delta_{6n^2}$}
Recalling that $n$ is even, we have
$\Delta_{6n^2} \cong (\IZ_n \times \IZ_n) \rtimes
S_3$ with $\IZ_n \times \IZ_n$ normal and thus we are once more aided
by Theorem \ref{sequence}.

We let $N := \IZ_n \times \IZ_n$ and $T := S_3$ and the exact sequence
assumes the form
\[
1 \rightarrow H^1(S_3,U) \rightarrow \tilde{M}(\Delta_{6n^2})
\rightarrow (\IZ_n)^{S_3} \rightarrow H^2(S_3,U)
\]
where $U := \Hom(\IZ_n \times \IZ_n,\IC^*)$ as defined in the previous
subsection.

By calculations entirely analogous to the case for $\Delta_{3n^2}$, we
have $(\IZ_n)^{S_3} \cong \IZ_2$. This is straight-forward to show.
Let $S_3 := \gen{z,w | z^3 = w^2 = \II, zw = wz^2}$. We see that it
contains $\IZ^3 = \gen{z|z^3=\II}$ as a subgroup, which we have
treated in the previous section. In addition to \eref{conj}, we have
\[
w^{-1}x^a y^b w = x^{-1-b}y^b = w x^a y^b w^{-1}.
\]
Using the form of the cocycle in Proposition \ref{zeta}, we see that 
$c_w(\alpha) = \alpha^{-1}$. Remembering that $c_z(\alpha) = \alpha$
from before, we see that the $S_3$-stable part of consists of
$\alpha^m$ with $m=0$ and $n/2$ (recall that in our case of
$\Delta(6n^2)$, $n$ is even), giving us a $\IZ_2$.

Moreover we have $H^1(S_3,U) \cong \II$. This is again easy to show.
In analogy to \eref{conjZ3}, we have
\[
w \cdot (p,q) = (-q, q-p), \quad\mbox{ for }(p,q) \in U,
\]
using which we find that $Z^1$ consists of $f : S_3 \rightarrow U$
given by $f(z) = (l_1,3k_2-l_1)$ and $f(w) = (2k_2,k_2)$. In addition
$B^1$ consists of $f(z) = (k-2l,-l-k)$ and $f(w) = (-2l,-l)$. Whence
we see instantly that $H^1$ is trivial.
 
Now in fact $H^2(S_3,U) \cong \II$ as well (the involved details of these
computations are too pathological to be even included in an appendix
and we have resisted the urge to write an appendix for the appendix). 

The exact sequence then forces immediately that
$\tilde{M}(\Delta_{6n^2}) \cong \IZ_2$. Moreover, since $M(S_3) \cong
\II$ (q.v. e.g. \cite{Karp}), by Part (i) of Theorem
\ref{sequence}, we conclude that
\[
M(\Delta_{6n^2}) \cong \IZ_2
\]
as reported in Table \eref{SU3}.
\newpage
\section{Appendix B: Intransitive subgroups of $SU(3)$}

The computation of the Schur Multipliers for the non-Abelian 
intransitive subgroups of $SU(3)$ involves some subtleties related to the
precise definition and construction of the groups.

Let us consider the case of combining the generators of $\IZ_n$ with
these of $\widehat{D_{2m}}$ to construct the intransitive subgroup
$<\IZ_n, \widehat{D_{2m}}>$. We can take the generators of
$\widehat{D_{2m}}$ to be 
\[
\label{matrix_1}
\alpha=\left( \begin{array}{ccc} \omega_{2m}  & 0 & 0 \\
0 & \omega_{2m}^{-1} & 0 \\ 0 & 0  & 1  \end{array} \right) ,~~~~
\beta=\left( \begin{array}{ccc} 0 & i & 0 \\ i & 0 & 0 \\ 
 0 & 0  & 1 \end{array} \right)
\]
and that of $\IZ_n$ to be
\[
\gamma=\left(  \begin{array}{ccc} \omega_n & 0 & 0 \\
0 & \omega_n & 0 \\ 0 & 0 & \omega_n^{-2}  \end{array} \right).
\]
 
The group $<\IZ_n, \widehat{D_{2m}}>$ is not in general the direct 
product of $\IZ_n$ and $\widehat{D_{2m}}$.
More specifically, when $n$ is odd $<\IZ_n, \widehat{D_{2m}}> = 
\IZ_n \times \widehat{D_{2m}}$. For $n$ even however, we notice that
$\alpha^m=\beta^2=\gamma^{n/2}$. Accordingly, we conclude that 
 $<\IZ_n, \widehat{D_{2m}}> = (\IZ_n \times \widehat{D_{2m}})/ \IZ_2$
for $n$ even where the central $\IZ_2$ is generated by $\gamma^{n/2}$.
Actually the conditions are more refined: 
when $n = 2(2k+1)$ we have $\IZ_n=\IZ_2 \times
\IZ_{2k+1}$ and so $(\IZ_2 \times \widehat{D_{2m}})/ \IZ_2 = 
\IZ_{2k+1} \times \widehat{D_{2m}}$.
Thus the only non-trivial case is when $n=4k$.

This subtlety in the group structure holds for all the cases where
$\IZ_n$ is combined with binary groups $\widehat{G}$. When $n~{\rm
mod}~4 \neq 0$, $<\IZ_n, \widehat{G}>$ is the direct product 
of $\widehat{G}$ with either $\IZ_n$ or $\IZ_{n/2}$. For $n~{\rm mod}~4 = 0$ 
it is the quotient group $(\IZ_n \times \widehat{G})/\IZ_2$.
In summary
\[
<\IZ_n, \widehat{G}> = \left\{\ba{lc} 
\IZ_n  \times \widehat{G} & n~{\rm mod}~2 = 1 \\
\IZ_{n/2}  \times \widehat{G} & n~{\rm mod}~4 = 2 \\
(\IZ_n \times \widehat{G})/\IZ_2  & n~{\rm mod}~4 = 0 
\ea \right. .
\]

The case of $\IZ_n$ combined with the ordinary dihedral group
$D_{2m}$ is a bit different however.
The matrix forms of the generators are 
\[
\label{matrix_2}
\alpha=\left( \begin{array}{ccc} \omega_{m}  & 0 & 0 \\
0 & \omega_{m}^{-1} & 0 \\ 0 & 0  & 1  \end{array} \right) ,~~~~
\beta=\left( \begin{array}{ccc} 0 & 1 & 0 \\ 1 & 0 & 0 \\ 
 0 & 0  & -1 \end{array} \right) ,~~~~
\gamma=\left( \begin{array}{ccc} \omega_n & 0 & 0 \\
0 & \omega_n & 0 \\ 0 & 0 & \omega_n^{-2}  \end{array} \right)
\]
where $\alpha$ and $\beta$ generate $D_{2m}$ and $\gamma$ generates
$\IZ_n$.

From these we notice that when both $n$ and $m$ are even,
$\alpha^{m/2}=\gamma^{n/2}$ and $<\IZ_n,D_{2m}>$ is not a direct
product. After inspection, we find that 
\[
<\IZ_n, D_{2m}> = \left\{\ba{lc} 
\IZ_n  \times D_{2m} & m~{\rm mod}~2 = 1 \\
\IZ_n  \times D_{2m} & m~{\rm mod}~2 = 0, n~{\rm mod}~2 = 1  \\
\IZ_{n/2} \times D_{2m} &  m~{\rm mod}~2 = 0, n~{\rm mod}~4 = 2 \\
(\IZ_n \times D_{2m})/\IZ_2  &  m~{\rm mod}~2 = 0, n~{\rm mod}~4 = 0
\ea \right. .
\]
The Schur Multipliers of the 
direct product cases are immediately computable by consulting Theorem
\ref{directprod}. For example,
$M(\IZ_n \times \widehat{D_{2m}}) \cong M(\IZ_n) \times M(\widehat{D_{2m}})
\times (\IZ_n \otimes \widehat{D_{2m}})$ by Theorem
\ref{directprod}, the last term of
which in turn equates to
$\Hom(\IZ_n,\widehat{D_{2m}}/\widehat{D_{2m}}')$. This is 
$\Hom(\IZ_n,\IZ_2 \times \IZ_2) \cong \IZ_{\gcd(n,2)} \times
\IZ_{\gcd(n,2)}$ for $m$ even and 
$\Hom(\IZ_n,\IZ_4) \cong \IZ_{\gcd(n,4)}$ for $m$ odd.
By similar token, we have that $M(\IZ_n \times D_{2m})$ for
even $m$ is $\IZ_2 \times \Hom(\IZ_n,\IZ_2 \times \IZ_2) \cong
\IZ_2 \times \IZ_{\gcd(n,2)} \times \IZ_{\gcd(n,2)}$
and $\Hom(\IZ_n,\IZ_2) \cong \IZ_{\gcd(n,2)}$ 
for odd $m$. Likewise $M(\IZ_n \times \widehat{E_{6,7,8}}) =
\Hom(\IZ_n,\IZ_{3,2,1})$.
\bibliographystyle{JHEP}

\end{document}